\documentclass[aps,pre,twocolumn,showpacs,amsmath,amssymb]{revtex4}
\usepackage{bm}
\usepackage{graphicx}

\newcommand{\be}{\begin{eqnarray}}
\newcommand{\ee}{\end{eqnarray}}

\newcommand{\ba}{\begin{array}}
\newcommand{\ea}{\end{array}}

\newcommand{\ket}[1]{|#1\rangle}

\begin{document}

\title{
On Entanglement with Vacuum }
\author{Marcin Paw{\l}owski and Marek Czachor}

\address{
Katedra Fizyki Teoretycznej i Metod Matematycznych\\
Politechnika Gda\'nska, 80-952 Gda\'nsk, Poland}

\begin{abstract}
The so-called entanglement with vacuum is not a property of the Fock space, but of some rather pathological representations of CCR/CAR algebras. In some other Fock space representations the notion simply does not exist. We have checked all the main Gedanken experiments where the notion of entanglement with vacuum was used, and found that all the calculations could be performed at a representation-independent level. In particular any such experiment can be formulated in a Fock-space representation where the notion of entanglement with vacuum is meaningless. So, for the moment there is no single experiment where the notion is needed, and probably it is simply unphysical. 
\end{abstract}


\maketitle

\section{Introduction}
The word ``entanglement" has been introduced by Schroedinger in 1935
\cite{Ent1}, and the phenomenon has been an object of study ever
since (and even before - take the famous ``EPR paper" \cite{EPR},
published a few months earlier, for example). More recently the
concept of ``entanglement with vacuum" had been introduced to study
nonlocalities of single-particle states \cite{Hardy,TWC}. Though it
had been strongly criticized in \cite{GHZ} and later in \cite{VW1},
this concept has been exploited to develop Quantum Key Distribution
protocols \cite{QKDE1,QKDE2} and further study of single-particle
nonlocality \cite{QTE,EWV}. Since quantum cryptography is nowadays
on the verge of becoming an applied science there is an obvious need
to clarify the issues concerning entanglement with vacuum. The aim
of this paper is to provide such clarification.

\section{Representations of CCR/CAR algebras}

In \cite{GHZ} authors point out that some states that seem to be
entangled in Fock space are merely single-particle ones in configuration
space and vice versa. On the other hand, authors of \cite{EWV} state
that: ``since the Fock basis is a complete basis, it is just as good
as any other to express and calculate quantum physics". The Fock
space mentioned in both papers is the first thing that needs
clarification.

The one particle-vacuum entangled state in that space is of the
form:
\be
\ket{\psi}=\frac{1}{\sqrt{2}}\left(\ket{1}_A\ket{0}_B + \ket{0}_A\ket{1}_B\right)
\ee
Indices $A$ and $B$ denote here modes occupied by the particle. If we would like to be more precise,
we should rather write it as:
\be
\ket{\Psi}=
\ldots\ket{0}_{i_1}\ket{0}_{i_2}\ket{0}_{i_3}
\ket{\psi}
\ket{0}_{j_1}\ket{0}_{j_2}\ket{0}_{j_3}\ldots
\ee
Other vacua represent here other modes that are not taken into account during while discussing the experiment.
A ``total vacuum", that is the state with no particles in any of the modes, can be put down as follows:
\be
\ket{{\bf 0}}=\ldots\ket{0}\ket{0}\ket{0}\ket{0}\ket{0}\ket{0}\ket{0}\ldots
\ee
State $\ket{\Psi}$ can also be expressed in second quantization formalism as:
\be \label{a+b}
\ket{\Psi}=\frac{1}{\sqrt{2}}\left(a_A^\dag +a_B^\dag \right)\ket{{\bf 0}}
\ee
In this, more general, approach entanglement does clearly not exist. It appears only when a specific representation
of the CAR or CCR algebra is chosen. If we choose the representation discussed above (we will call it MVR for Multiple
Vacua Representation, because the state $\ket{{\bf 0}}$ consists of infinite, and even uncountable, number of empty
modes --- vacua) the entanglement with vacuum appears. But we can choose another representation of those algebras, like
the one proposed by Berezin \cite{Ber}. Obviously (\ref{a+b}) does not change, but the explicit form of creation operators and vacuum does. If we use that representation then the ``total vacuum" is a vector:
\be
\ket{{\bf 0}}=\left(
\begin{array}{c}
1  \\
0  \\
0  \\
.  \\
.  \\
.  \\
\end{array}
\right)
\ee
and this is the only vacuum there, so there can be no entanglement with it (entanglement with a single vector is trivial). Here $\ket{\Psi}$ is also a vector from the Fock space but in this space entanglement with vacuum is not visible.

In \cite{GHZ} authors noticed that the entanglement with vacuum does not appear in second quantization formalism. This statement is imprecise. What we have just shown is that it  appears in the Fock space only if a specific representation of CCR or CAR algebra is chosen. It seems that rather than being a physical phenomenon the entanglement with vacuum is only a peculiarity of one of the infinite number of possible representations. In the following section we will try to judge if that is the case.

\section{The case study}
First of all it is worth noticing that MVR and Berezin's representation are inequivalent. The fact that the Fock space corresponding to the second one is separable and to the first one is not is enough to prove that. Since entanglement with vacuum appears only in one of them, there are four possibilities:

a) There is a preferred representation of the universe. It is MVR (or similar) and entanglement with vacuum exists and can be exploited.

b) There is a preferred representation of the universe, but it is Berezin's (or similar) and entanglement with vacuum is a meaningless concept.

c) Both representations are correct but they describe different entities.

d) All irreducible representations are physically equivalent \cite{foot} and there is no experiment that can be conducted to decide whether entanglement with vacuum has physical meaning or whether it is just a convenient notion for expressing more complex ideas. 

Natural method for finding which of these is true is to repeat calculations done in MVR but this time using another representation or representation independent formalism and compare results.
In papers \cite{TWC},\cite{QKDE1} - \cite{EWV} single photon's presence is being felt at two spatially separated phase sensitive
detectors. By phase sensitive detector we mean here the detection unit which consists of a pair of detectors,
beam splitter and light source, such as  the homodyne detectors in \cite{TWC}. The description of the beam splitter
which is  the only component of experiments described explicitlyin those papers, slightly
differs but this does not play a significant role. The unitary transformation performed by the beam splitter is
described by $B_1$ in \cite{TWC} and $B_2$ in \cite{QKDE1,QKDE2,QTE,EWV}:
\be
B_1=\frac{1}{\sqrt{2}}  \left(
\begin{array}{cc}
1 & i \\
i & 1 \\
\end{array}
\right)
\quad
B_2=\frac{1}{\sqrt{2}}  \left(
\begin{array}{cc}
1 & -1 \\
1 & 1 \\
\end{array}
\right)
\ee
The symmetric beam splitter $B_1$ might be realized by two back-to-back prisms with an air gap between, while the
antisymmetric one $B_2$ by silvering a glass plate on one side \cite{Yurke}. Evolution operators corresponding to those
matrices are, respectively:
\be
S_1=e^{ \frac{1}{4}i\pi \left( a_1^\dag a_2 + a_2^\dag a_1 \right)}
\quad
S_2=e^{ \frac{1}{2}\pi \left( a_1^\dag a_2 - a_2^\dag a_1 \right)}
\ee
satisfying:
\be
S_i^\dag \vec{a}^{\ out} S_i=B_i \vec{a}^{\ in}
\ee
In \cite{Hardy} particles in experiment are fermions, but it does not make much difference. To make calculations easier it is worth noticing that any state transformation described by 2x2 unitary matrix leads to the evolution operator of the form:
\be
S=e^K \quad K=\sum_{i,j}c_{i,j}a_i^\dag a_j
\ee
When considering the action of $S$ upon any state it is only commutator $\left[K,a_j^\dag\right]$ that matters. It is
straightforward to check that
\be \label{comm}
\left[K,a_j^\dag\right]=\sum_{i}c_{i,j}a_i^\dag
\ee
is true regardless whether annihilation/creation operators correspond to bosons and follow CCR or fermions and follow CAR.

Having the $S$ operators it is elementary to find the outcome of any of the experiments described in papers \cite{TWC},\cite{QKDE1} - \cite{EWV} using only CCR and the fact that $a\ket{{\bf 0}}=0$ (operators corresponding to
detectors are, of course, given by $N=a^\dag a$).

In \cite{Hardy} the matrix operator for the electron-positron annihilation point can not be given since in the number basis this operation is nonlinear. But we can introduce a $S$ operator for this point of the form:
\be
S_a=e^{\theta(a^\dag bd + ab^\dag d^\dag ) }
\ee
where $\sin\theta$ plays the role of probability amplitude of electron-positron annihilation, $a$ corresponds to the
state of photon field after annihilation, $b$ and $d$ to electron and positron respectively. If we choose
$\theta=\frac{\pi}{2}$ which corresponds to the assumption made by Hardy that if positron and electron meet the
annihilation is certain, we can once again get all the results by using only CAR.

As an example of operator nonlinear in the number basis for bosons we can give the one corresponding to the Kerr medium as given in \cite{NH}. It is easy to check that the CNOT gate for the dual-rail representation of the qubit that employs Kerr medium, works as it should by taking operators corresponding to the beam splitters as:
\be
S_B=e^{\frac{1}{2}\pi \left( a_3^\dag(t) a_4(t) - a_3^\dag(t) a_4(t) \right)  }
\ee
Kerr medium:
\be
S_K=e^{i\pi a_1^\dag(t) a_1(t) a_3^\dag(t) a_3(t)}
\ee
and phase shifters:
\be
S_\pi=e^{i\pi a_4^\dag(t) a_4(t) }
\ee

\begin{center}
\includegraphics[scale=.4]{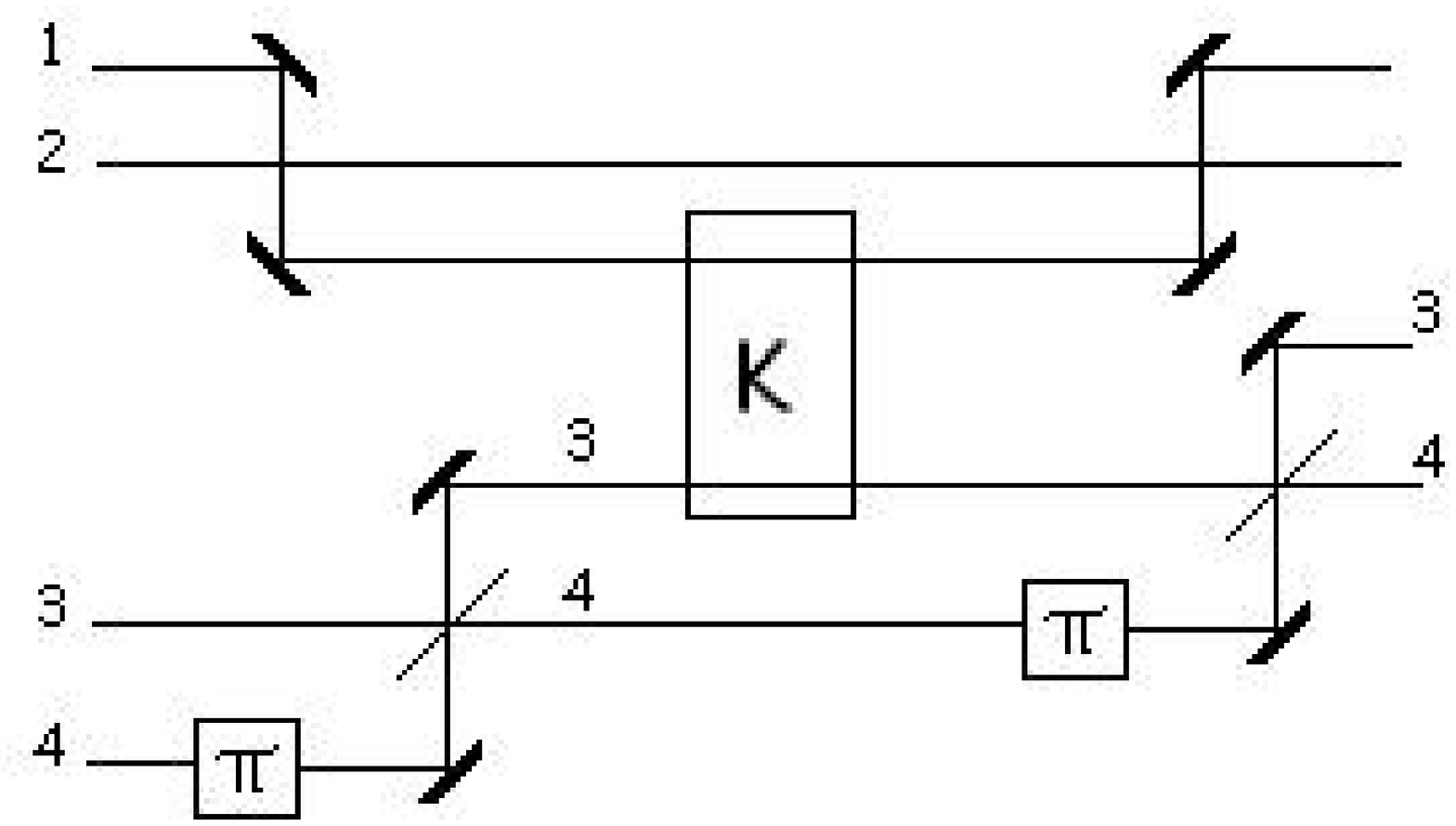}
{\small Fig 1. CNOT gate for dual-rail representation of the qubit}
\end{center}

Furthermore, since Kerr media, beam splitters and phase shifters are sufficient for quantum computation \cite{NH},  calculation of any quantum circuit will give the same results regardless of the representation chosen. And this means that the same physics follows from all the Fock representations --- including entanglement with vacuum or not --- and thus the presence of this questionable concept is only a question of taste.

\acknowledgments

This work is a part of the Polish Ministry of Scientific Research and Information Technology
(solicited) project PZB-MIN 008/P03/2003. MP would like to thank M. Kuna and M. \.Zukowski for
their help in accessing some of the following bibliography.


\begin{thebibliography}{99}
\bibitem{Ent1}E. Schroedinger, Naturwiss. 1935 {\bf 48}:807; {\bf 49}:823,844
\bibitem{EPR}A. Einstein, B. Podolsky and N. Rosen, Phys.Rev. {\bf 47}, 777 (1935)
\bibitem{Hardy}L. Hardy, {\it Quantum mechanics, local realistic theories, and Lorentz-invariant realistic theories},
Phys. Rev. Lett. {\bf 68}, 2981 (1992)
\bibitem{TWC}S.M. Tan, D.F. Walls, M.J. Collett, {\it Nonlocality of a single photon}, Phys. Rev. Lett. {\bf 66}, 252 (1991)
\bibitem{GHZ}D.M. Greenberger, M.A. Horne, A. Zeilinger {\it Tangled
Concepts about Entangled States}, Quantum Interferometry, F.
DeMartini et al. (Eds.), VCH Publishers, Weinheim (1996) 119
\bibitem{VW1} H.M. Wiseman, J.A. Vaccaro {\it The entanglement of
indistinguishable particles shared between two parties},
quant-ph/0210002
\bibitem{QKDE1}J.W. Lee, E.K. Lee, Y.W. Chung, H.W. Lee, J. Kim, {\it Quantum cryptography using single particle entanglement},
Phys. Rev. A {\bf 68}, 012324 (2003)
\bibitem{QKDE2}G.L. Giorgi, {\it Quantum Key Distribution with vacuum-one photon entangled states}, to appear in Phys. Rev. A,
quant-ph/0504150
\bibitem{QTE}H.W. Lee, J. Kim, {\it Quantum teleportation and Bell's inequality using single particle entanglement},
Phys. Rev. A {\bf 63}, 012305 (2000)
\bibitem{EWV}G. Bj\"{o}rk, P. Jonsson, L.L. S\'{a}nchez-Soto, {\it Single particle nonlocality and entanglement with vacuum}
Phys. Rev. A {\bf 64}, 042106 (2001)
\bibitem{Ber}F.A. Berezin, {\it Method of second quantization}, Academic Press, New York, 1966
\bibitem{foot}As shown in \cite{WC} all irreducible representations  occuring in the Jaynes-Cummings model are physically equivalent even though they are inequivalent in the sense of representation theory. Simultaneously, the reducible representations may lead to new physics. In the present paper we speak of irreducible representations.
\bibitem{Yurke}B. Yurke, {\it Squeezed Light}, (Rochester, 1989)
\bibitem{NH}M.A. Nielsen, I.L. Chuang, {\it Quantum computation and quantum infromation}, Cambridge University Press, (Cambridge, 2000)
\bibitem{SvN1}O. Bratteli, D.W. Robinson, {\it Operator Algebras and Quantum Statistical Mechanics}, Springer-Verlag, (New York, 1981)
\bibitem{SvN2}D. Petz, {\it An invitation to the algebra of Canonical Commutation Realations}, (Leuven University Press, 1990)
\bibitem{WC}M. Wilczewski, M. Czachor, quant-ph/0507093
\end{thebibliography}
\end{document}